# Navigating pollution: A multimodal approach to traffic and exposure management

Yueqi Liu[a], Ke Han[b*], Lei Yu[a], Wenrui Tu[a]

[a]School of Transportation and Logistics, Southwest Jiaotong University, Chengdu, China, 610000

[b]School of Economics and Management, Southwest Jiaotong University, Chengdu, China, 610000
[*]Corresponding author, e-mail: kehan@swjtu.edu.cn

**Abstract**

Few studies quantify how traffic management dynamically reshapes modal split and emission-exposure outcomes over pollution severities. This paper proposes a novel day-to-day assignment model integrating exposure cost, which includes exposure perception and emissions-dispersion-exposure algorithm. Numerical experiments reveal that and various levels of traffic-related measures have an air pollution scenario-dependent effect on the MT system. In light pollution scenarios, vehicle restrictions and reduced fares for buses or ridesharing help lower car usage and reduce emissions and exposure. However, under heavy pollution, higher-level restrictions and ridesharing fares paradoxically increase travelers' exposure by 18% and 6.3%, respectively, due to modal shift. Furthermore, timely pollution information updates could plausibly encourage healthier travel. This paper also proposes practical strategies for both routine and emergency traffic management, considering the trade-offs among travel cost, emission, and exposure, and emphasizes the need for measures tailored to different air pollution contexts to offer deeper insights for urban traffic policies.

---

1. **Introduction**

Exposure to air pollution has become a critical global issue (Boogaard et al., 2022), over 90% of the population is exposed to unhealthy levels of air pollution (Liang and Gong, 2020), contributing to millions of premature deaths annually (Vos et al., 2020). On the one hand, the transportation system significantly contributes to air pollutant emissions in urban areas (Xu et al., 2025). Moreover, travel exposure, which refers to the concentration of air pollutants that individuals encounter during their journeys, hereafter referred to as exposure, poses serious health risks (Zhang et al., 2023a, Engström and Forsberg, 2019, Khreis et al., 2017), especially during severe pollution periods.

The exposure levels wolud be influenced by travel time, pollutant concentrations in the surrounding environment, and the chosen traffic modal (Meena and Goswami, 2024, Guzman et al., 2023). Furthermore, exposure significantly influences travel behavior, prompting individuals to change their transportation mode, alter travel routes, or even cancel intineary (Wu et al., 2024, Zhao et al., 2018, Singh et al., 2021, Romero et al., 2019). Many researchers have noted that individuals with higher degree of perception of air pollution exposure are more likely to adjust their behavior (Zhang et al., 2023b, Dabirinejad et al., 2024).

Traffic policies and control measures, such as vehicle restrictions, congestion pricing, and public transport incentives, play a crucial role in mitigating urban air pollution under both routine conditions and emergency scenarios (Wu et al., 2022, Khreis et al., 2023, Han et al., 2024, de Buen Kalman, 2021, Webster, 2024). These measures aim to influence travel behavior to reduce emissions and, consequently, lower pollutant levels. This interrelationship between travel behavior, emissions, and exposure is dynamic and reciprocal. Changes in travel behavior can lead to variations in emissions and exposure levels, which, in turn, influence future travel behavior. Such feedback loops may alter the overall performance of multimodal transportation (MT) systems, potentially stabilizing at a new equilibrium. Simultainasly, the increasing complexity of transportation systems has introduced more challenges, such as emergence of new transportation modes (Sun et al., 2022, Bakirci, 2024), and ensuring effective responses to pollution (Cao, 2024, Chowdhury et al., 2017), for both routine and emergency traffic management aimed at mitigating air pollution exposure.

Addressing these challenges requires a comprehensive strategy that not only focuses on reducing emissions but also fully considers the impact of exposure on travelers. Although significant progress has been made in transportation modeling, existing studies primarily focus on optimizing time costs, reducing emissions, and lowering exposure (Ma et al., 2017, Long et al., 2018, Wang et al., 2018, Lu et al., 2016, Zhang et al., 2013, Bin Thaneya and Horvath, 2023). Moreover, most of these studies concentrate on single-mode transportation systems. However, solely minimizing emissions or exposure may not always be the most effective approach. As noted by Kickhöfer and Kern (2015), policy-making often requires a trade-off between reducing exposure and emissions.

"Furthermore, current models primarily examine the unidirectional impact of travel behavior or traffic management measures on exposure, as in Han et al. (2024), often overlooking the concomitant feedback effects of exposure dynamics on travelers' behavioral adaptations. This gap could lead to biased predictions regarding the dynamic evolution of MT systems. Therefore, it is crucial to integrate exposure factors into traffic assignment models, especially as exposure costs have been monetized based on healthcare expenses incurred (Alexeeff et al., 2022, Rodrigues et al., 2020), providing a potential solution. For example, Kickhöfer and Kern (2015) use exposure costs as an indicator for evaluating transportation policies. However, these studies have not considered the differential exposure across various transportation modes, nor have they explored the potential effects of exposure costs on transportation behavior.

To fill these gaps, this study proposes a MT model considering air pollution exposure, using numerical experiments, aiming to provide traffic management strategies for both routine pollution control and emergency air pollution responses. The main contributions of this study are as follows:

1. It integrates exposure costs into a Day-to-Day MT assignment model, combining it with emission factors, atmospheric dispersion, and exposure models, while considering exposure perception. This research explores the interaction between travel behavior, traffic emissions, and exposure, advancing MT modeling and providing theoretical support for optimizing related policies.

2. It analyzes the dynamic changes in MT system under heavy pollution, showing that after a heavy

pollution interference, the MT system undergoes abrupt changes before gradually stabilizing, leading to increased car use, emissions, and exposure.

3. It finds that traffic measures have different effects on MT system regarding air pollution exposure. Specifically, we compare the effects of vehicle restrictions, pricing (e.g., bus fares, ridesharing fares), and air quality information dissemination on modal choice, traffic emissions, and exposure in both light and heavy pollution scenarios.

4. It demonstrates that traffic measures have different effects depending on the pollution scenario. Finally, it proposes strategies for routine management and emergency management during heavy pollution. These recommendations offer guidance for urban management decisions, emphasizing that stressing that traffic policies must be context-specific to ensure effectiveness.

The remainder of this paper is organized as follows. The following section presents a brief literature review. Section 3 presents the methodology for the modeling of traffic assignment, traffic emssion, air dispersion, and travel exposure. Section 4 decribes the numerical experiments and results. Section 5 discusses policy implications on routine and emergency managemen. The final section concludes our findings and limitations.

## 2. Literature review

### 2.1. Routine and emergency traffic management

Concerning air pollution, traffic management can be divided into routine traffic management (RTM) and emergency traffic management (ETM), based on whether it addresses suddenly heavy pollution events. Regarding ETM, governments might issue air pollution alerts (Zhang et al., 2024) and implement stricter measures to rapidly reduce emissions and restore urban air quality, such as enhancing vehicle restriction, further reducing public transport fares, increasing tolls, or temporarily adjusting traffic organization (Huang et al., 2017, Han et al., 2024, Han et al., 2020, Rivera, 2021),. However, the effects of these measures are mixed and still warrant further exploration (Sanchez et al., 2020, Guerra et al., 2022). On the other hand, many studies primarily focus on changes in the external traffic environment, such as reduced traffic emissions or improved air quality (Bigazzi and Rouleau, 2017, Webster, 2024), but they tend to ignore the impact of transportation modal shifts on exposure. Moreover, during heavy pollution episodes, although ETM might improve air quality across the city, the shift in transportation modes could expose certain groups to higher risks (Han et al., 2024). Therefore, this paper further investigates the dual impacts of regular and emergency traffic management measures on both emissions and exposure, aiming to provide more robust support for policy development and decision-making.

*2.2. Travel exposure assessment*

Advances in air pollution monitoring technology and the application of mobile sensors have enabled researchers to collect air quality data and assess personalized exposure (Mishra et al., 2016, Li et al., 2022, Ji et al., 2023), highlighting the differential impacts of various transportation modes on traveler exposure (Singh et al., 2021). Numerous studies have demonstrated significant variations in exposure levels across transportation modes, particularly between private cars and public transit. Passengers of public transport are typically exposed to higher concentrations of pollutants, due to both the air quality inside the vehicle and external travel process, such as walking to and waiting at stations (Jing et al., 2017, Tran et al., 2021, Qiu et al., 2017, Liu et al., 2024b). Some studies utilize data from roadside air monitoring stations and in-vehicle travel time to assess exposure levels across different transportation modes (Kousa et al., 2002), ignoring the in-vehicle air environment and travel process outside the vehicle. On this basis, (Liu et al., 2024b) further incorporate both in-vehicle and out-vehicle exposure into the assessment from the perspective of the entire trip process.

*2.3. Generalized travel cost calculation*

Generalized travel cost is a critical metric in transportation evaluation and serves as a core element in traffic assignment models. It typically includes monetized factors such as time, crowding, energy consumption, and fares, as in (Pi et al., 2019, Wei et al., 2020). The monetization of travel time is often achieved using a value of time (VOT) coefficient (Blayac and Causse, 2001, Zhao et al., 2016). Time cost is usually evaluated on a "door-to-door" basis, especially considering the walking and waiting time for public transport (Li et al., 2023, Kawakami and Shi, 1994). With the growing attention on air pollution exposure, an increasing number of studies have started to consider its health impacts, along with the associated health costs (Alexeeff et al., 2022). Kickhöfer and Kern (2015) applies exposure costs in transportation policy assessments to optimize transportation strategies.

*2.4. Integration of traffic, emission, and exposure models*

The intricate relationship between travel behavior, emissions, and exposure necessitates the integration of traffic, emission, and exposure models (Singh et al., 2021, Meena and Goswami, 2024). Most existing studies on traffic emissions and exposure analysis cascade three types of models. Initially, traffic model outputs, such as flow and speed, serve as inputs to the emission model, which then calculates exposure levels. For example, Zhang et al. (2013) analyze CO exposure at intersections with different signal timing schemes using a traffic cell-transmission model, emission factors, and a Gaussian plume dispersion model. Vallamsundar et al. (2016) use simulation software to connect the Light-weight Dynamic Traffic Assignment Engine (DTALite), MOVES emissions model, and AERMOD air dispersion model to assess $PM_{2.5}$ exposure in large urban settings. However, it is insufficient to examine emissions and exposure alone. Because traffic behavior not only influences exposure but is also affected by it in a feedback loop, which requires particular attention, especially when responding to traffic policies (Bigazzi and Rouleau, 2017). This paper proposes integrating exposure costs into the day-to-day (DTD) traffic assignment model to explore the dynamic feedback between behavior, emissions, and exposure.

3. **Method**

The day-to-day traffic assignment model is widely used to predict dynamic changes in transportation systems, offering a theoretical foundation for traffic management, policy-making, and system optimization(Yu et al., 2020, Liu et al., 2024a, Zhu et al., 2019). The model examined in this study is based on the multi-modal day-to-day traffic assignment framework proposed by Liu et al. (2024c), incorporating two types of travelers (car owners and non-car owners), three transportation vehicles (private cars, buses, and subways), and five travel choices (solo driving, ridesharing driving, ridesharing riding, bus riding, and subway riding). This model can be applied to evaluate the effects of traffic policies on multi-modal transportation systems.

To account for travelers' perception of exposure risks and their behaviour to reduce them (Ajayi et al., 2023, Tribby et al., 2013), we extend the model by incorporating exposure cost. The extended model integrates traffic emissions, air dispersion, background concentrations, and mode-specific exposure calculations, providing a foundation for experimental analysis of traffic management measures and air pollution disturbances.

*3.1. Generalized cost with exposure*

To assess the impact of exposure on mode and route choice in MT system, and its effect on overall system performance, this study incorporates exposure cost into the traditional generalized cost framework. Specifically, we introduce an exposure perception coefficient, which reflects travelers' awareness of exposure. This perception of air pollution influences travel behavior (Ma et al., 2021, Meena et al., 2024, Dabirinejad et al., 2024), with higher awareness leading travelers to adjust their travel mode to reduce exposure.

The exposure cost for each travel mode $EC_{r,m}^{w,g}$ is expressed as:

$$EC_{r,m}^{w,g} = EX_{r,m}^{w,g} \omega \eta_{ex} \quad (3.1)$$

in which $w$ is the origin-destination (OD) pair. $g$ is the traveler type. $m, r$ are the travel mode and route, respectively. $\omega \in [0,1]$ represents exposure perception, with higher values indicating stronger awareness; when $\omega = 0$, the traveler disregards exposure. $\eta_{ex}$ is the exposure value coefficient, which converts exposure to a cost equivalent. $EX_{r,m}$ is the exposure level for a given mode $m$ and route $r$.

The generalized cost for each mode, excluding exposure costs, is detailed in (Liu et al., 2024c). For simplicity, we denote this as $C_{r,m}^{w,g}$. Thus, the total generalized cost for each mode, $C_{r,m}^{w,g\prime}$, is expanded as:

$$C_{r,m}^{w,g\prime} = C_{r,m}^{w,g} + EC_{r,m}^{w,g} \quad (3.2)$$

*3.2. Vehicle emission calculation*

Vehicle emissions release particulate matter of various sizes, which pose a serious threat to public health (Xu et al., 2025). This study adopts the approach of Zou et al. (2023) and Zhao et al. (2016), calculating vehicle PM$_{2.5}$ emissions based on the emission factor:

$$E_a = \sum_m EF_m \times f_{m,a} \times d_a \tag{3.3}$$

where $E_a$ represents the emission of air pollutants PM$_{2.5}$ of the link $a$, respectively (unit: g/h). $f_{m,a}$ is the traffic flow of mode $m$ of the link $a$ (unit: veh/h). $d_a$ is the distance of link $a$ ((unit: km). $EF_m$ denotes the emission factors of PM$_{2.5}$ for the vehicle of mode $m$ (unit: g/km/veh).

### 3.3. Air pollutants Dispersion

The vehicular-emitted air pollutants near roadways can be calculated by an air dispersion model (Schindler et al., 2021). Specifically, vehicles traveling on streets generate intermittent but stable pollutant concentration fields along the road, allowing emissions from each road to be treated as a finite line source (Luhar and Patil, 1989, Mishra et al., 2016). For a bidirectional road, the source strength of the finite line source $b$ is related to the sum of the unit emissions of bidirectional traffic and is expressed as:

$$Q_b = \frac{(E_{a^+} + E_{a^-}) \times 10^6}{L \times 3600} \tag{3.4}$$

where $Q_b$ is the source strength (unit: µg/m/s), indicating the traffic emission rate per unit distance. $E_{a^+}, E_{a^-}$ are the emissions (unit: g/h) for the bidirectional links on the line source $b$. $L$ is the length of the source $b$.

We utilize the General Finite Line Source Model (GFLSM) to analyze the diffusion concentration of PM$_{2.5}$ from vehicle emissions, which is widely used in road traffic emission dispersion studies due to its simple calculation method and ability to adapt to any wind direction (Luhar and Patil, 1989, Pilla and Broderick, 2015, Mishra et al., 2016). In the GFLSM model, the contribution of line source $b$ to the pollutant concentration at a receptor located at point $(x, y, z)$ is given by:

$$\begin{aligned}D'_b(x,y,z) = &\frac{Q_b}{2\sqrt{2\pi}\sigma_z u \times \sin\varphi}\left[\exp\left(\frac{(z-H)^2}{-2\sigma_z^2}\right) + \exp\left(\frac{(z+H)^2}{-2\sigma_z^2}\right)\right]\\ &\times \left[\text{erf}\left(\frac{(\sin\varphi(L/2-y)) - x\cos\varphi}{\sqrt{2}\sigma_y}\right) + \text{erf}\left(\frac{(\sin\varphi(L/2+y)) + x\cos\varphi}{\sqrt{2}\sigma_y}\right)\right]\end{aligned} \tag{3.5}$$

where receptor point $(x, y, z)$ is located in a coordinate system with the origin at the midpoint of line source $b$ (with the line source direction along the y-axis and the perpendicular direction along the x-axis) and is used to measure the downwind pollutant concentration $D_b'$. $x$ is downwind distance. $u$ is the average wind speed. $\varphi \in (0°, 180°)$ is the angle between the wind vector and the line source. $H$ is the height of the line source (which can be set to 0 for near-ground traffic emission measurements, with receptor height $z = 0$ (Tan et al., 2021). erf( ) is the Gauss error function. $\sigma_y, \sigma_z$ are the horizontal and vertical diffusion coefficients, respectively, represented by the Briggs (Hanna et al., 1982) dispersion coefficient (where $x \in (10^2, 10^4)$) based on Pasquill (Pasquill, 1961) stability class in urban areas (Pilla and Broderick, 2015).

### 3.4. Background concentration

The daily mean background concentration levels are influenced by the diffusion of pollutants from traffic source and the other sources (Pilla and Broderick, 2015). For simplicity, this study assumes the diffusion from the other sources is constant. It is assumed that the background concentration has an initial value, $D^0$, which represents the initial state of air pollution concentration in the MT system and characterizes different pollution scenarios. Therefore, considering the diffusion of vehicle emissions, the background concentration $D_b$ of line source $b$ is expressed as:

$$D_b = D_b' + D_b^0 \tag{3.6}$$

Therefore, the background concentration at the receptor point $(x_b, 0, 0)$ is the sum of the diffusion concentration $D_b'(x_b, 0, 0)$ and the initial air pollutant concentration $D^0$, as shown below:

$$D_b(x_b, 0,0) = \frac{Q_b}{\sqrt{2\pi}\sigma_z u \times \sin\varphi} \\ \times \left[ \mathrm{erf}\left(\frac{(\sin\varphi(L/2)) - x_b \cos\varphi}{\sqrt{2}\sigma_y}\right) + \mathrm{erf}\left(\frac{(\sin\varphi(L/2)) + x_b \cos\varphi}{\sqrt{2}\sigma_y}\right) \right] + D^0 \tag{3.7}$$

### 3.5. Exposure calculation

Travel exposure consists of two parts: inside-vehicle exposure $in\_EX_{m,r}^{w,g}$ and outside-vehicle exposure $out\_EX_{m,r}^{w,g}$ (Liu et al., 2024b). The travel exposure $EX_{r,m}^{w,g}$ for mode $m$ on route $r$ can be represented as:

$$EX_{m,r}^{w,g} = in\_EX_{m,r}^{w,g} + out\_EX_{m,r}^{w,g} \qquad (3.8)$$

*3.5.1. Inside-vehicle exposure*

In-vehicle exposure refers to the exposure experienced by travelers while using a mode of transportation. It is influenced by factors such as travel time, background concentration, and the vehicle's input/output (I/O) ratio (Vallamsundar et al., 2016). When a traveler passes through link $a$, the background concentration he or she exposed:

$$D_a = \delta_{b,a} D_b \quad (a \in r) \qquad (3.9)$$

where $\delta_{b,a}$ indicates whether link $a$ lies on line source $b$ (1 if true, 0 if false).

Therefore, for $h_{r,m}^{w,g}$ travelers using mode $m$ on route $r$, the in-vehicle exposure is expressed as:

$$in\_EX_{m,r}^{w,g} = \sum_{a \in r} t_a \times D_a \times I_m \times h_{r,m}^{w,g} \qquad (3.10)$$

where $t_a$ is the travel time passing through link $a$. $I_m$ is the I/O ratio of mode $m$.

*3.5.2. Out-of-vehicle exposure*

We consider that public transport passengers suffer outside-vehicle exposure while waiting for transit and walking to/from stations along the road. It assume that waiting time is half the vehicle departure frequency (Wei et al., 2020), and walking time is related to the service area of the station and the walking speed (Liu et al., 2024b). Compared to public transportation, solo and ridesharing trips are door-to-door, and their out-of-vehicle travel is typically omitted (Wei et al., 2020, Kawakami and Shi, 1994). Hence, the out-of-vehicle exposure for different modes is expressed as:

$$out\_EX_{m,r}^{w,g} = \begin{cases} \left(\dfrac{1}{2\rho} + t_{walk\_as}\right) D_{as} + t_{walk\_at} D_{at} & (as, at \in r), \text{for bus and metro} \\ 0, \text{for solo and ridesharing} \end{cases} \qquad (3.11)$$

where $D_{as}, D_{at}$ represent the pollutant concentrations on the start and terminal stops located at link $as$ and link $at$, respectively. $t_{walk\_as}, t_{walk\_at}$ are the time spent on walking to/from stations along roads, respectively.

Moreover, the total exposure of the travelers in MT system can be expressed:

$$EX = \sum_w \sum_g \sum_m \sum_r (in\_EX_{m,r}^{w,g} + out\_EX_{m,r}^{w,g}) \tag{3.12}$$

## 4. Numerical experiments

This study considers two scenarios: a "light pollution scenario (LPS)," where pollution is mainly caused by traffic emissions with a low initial concentration, and a "heavy pollution scenario (HPS)," characterized by a higher initial concentration due to a sudden increase in the light pollution levels. This section analyzes the dynamic changes in the MT system under varying pollution backgrounds, traffic measures, and exposure perceptions, focusing on changes in modal choice, travel costs, emissions, and exposure.

We use the Sioux-Falls road network as an example, which consists of 24 nodes, 76 links, 24 zones, and 38 finite sources, shown in Figure 1. We set the exposure value coefficient $\eta_{ex} = 0.1$ (Du et al., 2021) and examine the impact of varying exposure perception $\omega$. Additionally, studies show that $PM_{2.5}$ concentrations remain relatively high (around 80%) even at distances far from the pollution source (e.g., 400 meters) (Karner et al., 2010). Thus, we take a receptor at $x_b = 150$m. besides, it assumes a wind speed $u = 1$m/s and a wind direction angle $\varphi = \frac{\pi}{6}$.

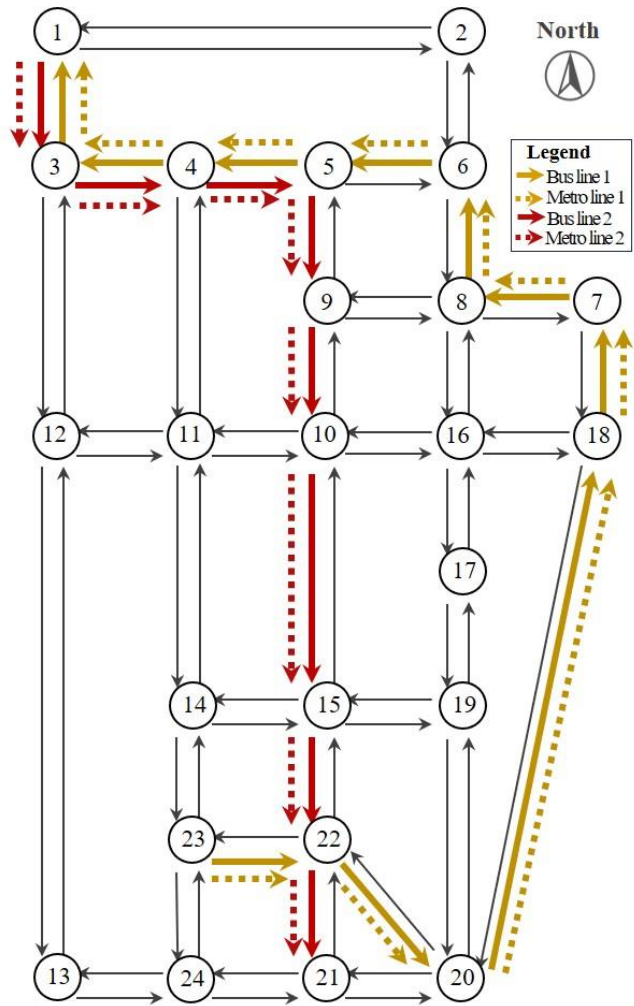

**Figure 1　Sioux-Falls network with public transport**

*4.1. System performance considering heavy pollution*

An increase in the initial background concentration raises air pollution in Day 16, shifting the system from a light to a heavy pollution condition. As shown in Figure 2, the system fluctuates significantly due to disturbances but eventually stabilize. Particularly, while the system would adapt and adjust in the short term, the worsening air pollution leads to increased car usage at last, posing challenges for environmental protection and raising health risks for travelers.

Figure 2(a) shows that heavy pollution suppresses public transportation demand. After the system stabilizes, there is a shift from public transit (bus, metro) to private cars (solo, ridesharing), with both solo and ridesharing trips increasing. Consequently, the Passenger Car Units (PCU) on the roads also rise. In Figure 2(b), the trends in in-vehicle time and travel time are opposite: in-vehicle time increases under HPS, as travelers prefer private cars or ridesharing, reducing their reliance on public transport. Conversely, travel time decreases, likely because travelers opt for more direct routes (e.g., solo or ridesharing), avoiding wait times and transfers associated with public transport. Additionally, the overall travelers' monetary cost rises. Furthermore, Figure 2(c) and Figure 2(d) show both system emissions and exposure levels worsen. Under HPS, exposure levels rise, increasing health risks for travelers. Emissions also increase, further degrading air quality and hindering improvements in travel conditions.

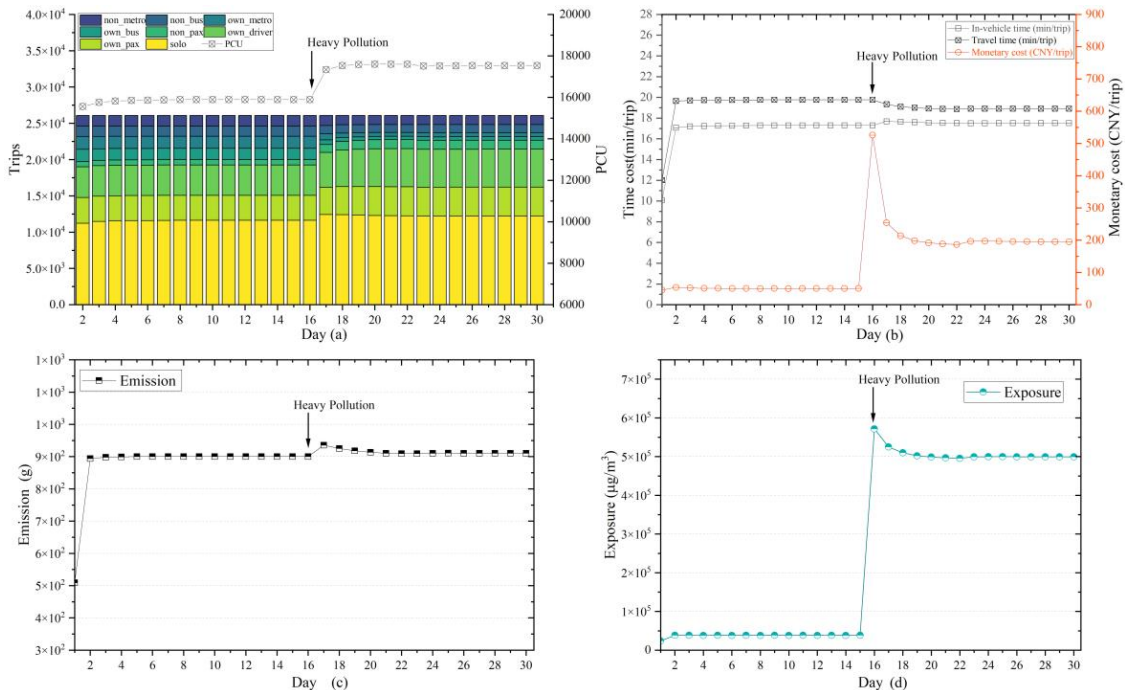

Figure 2　System changes under the increase in air pollution in Day 16

*4.2.　. Impact of different traffic measures*

This section analyzes the effects of various traffic management measures—such as restrictions of cars, changes in bus fare and ridesharing fare—on the MT system under LPS and HPS, by comparing changes in system performance under the steady state. Specifically, the study examines a range of restriction levels, from 0 to 0.9, representing progressively stricter measures. To ensure the availability of alternative travel options, the analysis is limited to OD pairs with public transit service. Additionally, the impact of bus fare changes is considered within a range of 1 to 10 CNY, with the public service considerations. Finally, ridesharing fare are examined within a range of 1 to 10 CNY per kilometer.

*4.2.1. Modal split, shift and modal exposure*

**1. Vehicle restriction**

As shown in Figure 3(a) and Figure 3(e), the increase in the intensity of restriction leads to a shift in travel modes, predominantly from private car use to public transport (bus and metro), regardless of whether the background pollution is light or heavy. To be precise:

- **Solo travel gradually decreases**: As the restriction ratio increases, particularly when it exceeds 0.5, the number of solo travelers decreases significantly, indicating that the restriction policy has a notable impact in reducing solo travel. In HPS, solo travel is more significantly reduced. For example, when the restriction ratio is 0.5, solo travel decreases from 12,222 to 11,147 under HPS, a reduction of 8.8%, which is greater than the 8.3% reduction under LPS. Ultimately, when the restriction ratio reaches 0.9, the number of solo travelers in HPS even falls below that of LPS.

- **Ridesharing gradually decreases**: ridesharing trips decreases with the increase of restriction intensity. While car restriction may increase the potential demand for ridesharing, the reduced availability of vehicles results in a decrease in ridesharing trips rather than an increase. Furthermore, under different restriction rates, ridesharing travel is consistently more popular in high pollution conditions than in low pollution conditions, although the magnitude of the decline is more significant in the former.

- **Public transport demand increases**: As the restriction ratio increases, limiting the supply of private cars, public transport (buses and metros) becomes the dominant alternative mode of travel. Notably, when the restriction ratio reaches 0.9, public transport accounts for more than 40% of travel demand.

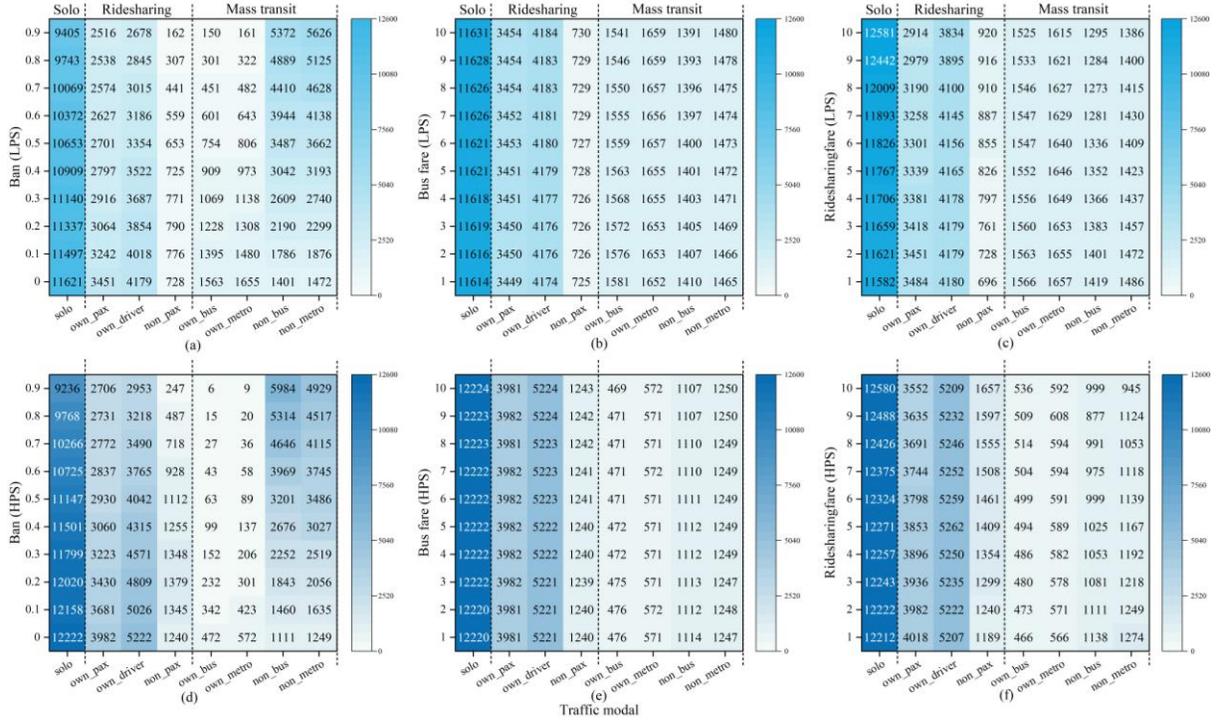

**Figure 3  Trips of each modal under different disturbances**

Meanwhile, different modal exposure of each transport modes is shown in Figure 4, which is positively correlated with modal split. The trend of modal exposure aligns with the trend in travel demand. Figure 5 presents the changes in average exposure for each mode and the system's average exposure. The modal exposure trends are as follows:

- As restrictions intensify, the exposure share of public transport gradually increases, with a more significant rise under the HPS. As shown in Figure 4(a) and Figure 4(e), the exposure share of public transport rises from 42% under the LPS to 58%, while under the HPS, it rises from 24% to 58%. This is also reflected in the comparison of average exposures across modes, as shown in Figure 5, where the exposure per trip for bus or metro users exceeds that for car users.

- Comparing  Figure 5(a) and  Figure 5(b), although vehicle restrictions reduce car use, exposure for all modes is more severe under HPS, particularly for public transport, where exposure increases significantly. This suggests that public transport becomes the main mode under more severe pollution, but its exposure risk also increases.

- In LPS, car restrictions reduce both system-wide and mode-specific per capita exposure, as shown in

Figure 5(a). In this context, the restriction measures effectively decrease exposure for travelers in all modes. However, in HPS, when both traffic restrictions and air pollution take effect together, the per capita exposure for the entire system gradually increases, as shown in Figure 5(b). This is because, under higher pollution levels, restricting private car use results in more travelers switching to public transport, leading to a larger increase in modal exposure and a rise in overall exposure. Thus, car restriction policies are more effective at reducing travel exposure when air quality is relatively good.

**2. Pricing measure--bus fare**

As shown in Figure 3(b) and Figure 3(e), with the increase in bus fares (from 1 CNY to 10 CNY), the changes in modal choices keep relatively stable, with only minor fluctuations. In particular:

- Under LPS, solo and ridesharing trips slightly increase, while in HPS, changes are almost negligible.
- With the increase in bus fares, the competitiveness of metro travel increases, causing a small shift in demand from buses to metro, but overall, the demand for both bus and metro remains relatively stable.

The impact of bus fare changes on travel modal choice is minimal for two main reasons: On the one hand, due to the public service nature of buses, bus fares are typically low and have a limited range of variation. On the other hand, other costs also count.

No matter in LPS or HPS, exposure of solo travel remains dominant among all modes, consistently at or above 45%. The exposure share of each modal remains stable with the rising fare, as shown in Figure 4(b) and Figure 4(e). Meanwhile, Figure 4(b) and Figure 5(e) depict that, bus fare changes have little effect on exposure. This suggests that adjusting bus fares alone is insufficient to reduce travel exposure. More comprehensive management measures are required to effectively protect public health during travel.

**3. Pricing measure--ridesharing fare**

Figure 3(c) and Figure 3(f) display that an increase in ridesharing fares leads to a rise in solo demand, without substitution of ridesharing by public transport. Particularly:

- Under LPS, as ridesharing fares increase, solo trips rise, while ridesharing trips decline. It is noteworthy that public transport usage also gradually decreases. This suggests that the rise in ridesharing fares suppresses ridesharing demand (both drivers and passengers), leading some travelers to switch to solo travel rather than public transport, and reducing public transport demand. The decreased attractiveness of public transport is due to the increase in cars on the road, which in turn leads to higher emissions and travel times. Considering exposure, some public transport travelers also switch to solo, which is why solo travel continues to rise.

- Under HPS, similarly, solo travel continues to rise, and public transport use continues to decline. However, due to the exacerbation of air pollution, the increase in ridesharing fares stimulates some demand for ridesharing, with more drivers joining to meet the demand of passengers seeking to avoid exposure risks (e.g., passengers who would otherwise take public transport). Consequently, ridesharing volumes initially increase slowly as the fare rises from 1 CNY/km to 9 CNY/km, before eventually decreasing.

as shown in Figure 4(c) and Figure 4(f), with the rise in ridesharing fares, solo travel exposure stays dominant, without falling below 45% under the LPS and 56% under HPS. Meanwhile, exposure to solo travel increases with higher fares, while the exposure share for public transport slightly decreases, and the exposure share for ridesharing remains mostly unchanged. Furthermore, as shown in Figure 5(c), under LPS, as ridesharing fares increase, both system-wide and mode-specific average exposures gradually rise. Under HPS, as shown in Figure 5(f), at higher fares (≥7 CNY/km), metro and bus exposures experience obvious fluctuations. This indicates that the combined impact of worsening air pollution and changes in ridesharing fares should not be overlooked, and a holistic approach is required to address them.

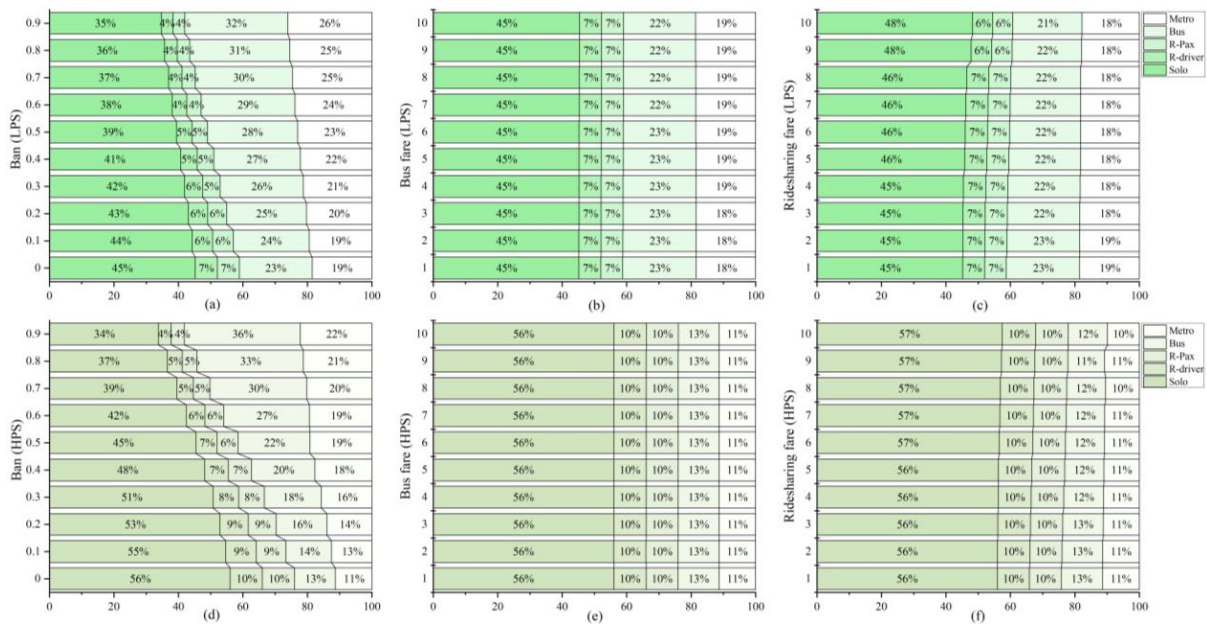

**Figure 4  Precentage share of each modal exposure under different disturbances**

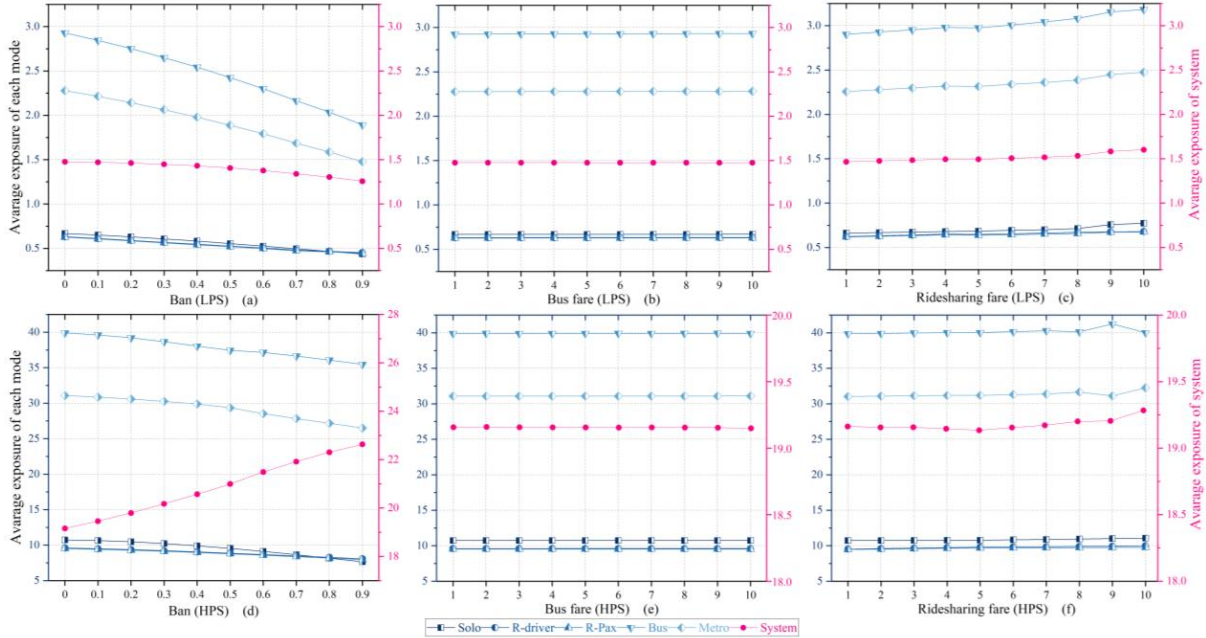

**Figure 5  System-wide and mode-specific average exposures under different measures**

*4.2.2. System emissions and exposure*

System exposure reflects the current exposure situation of travelers. Moreover, changes in system emissions are an important indicator for future exposure changes. When system emissions are reduced, the air quality will improve, which in turn contributes to safeguarding the health of urban residents. Under different traffic measures, the changes in system $PM_{2.5}$ emissions and exposure are as follows:

1. **Under vehicle restriction**

As shown in Figure 6(b), under HPS, vehicle restrictions increase system exposure by 2%-18%. However, at the same time, $PM_{2.5}$ emissions decrease significantly. Specifically, when the restriction ratio exceeds 0.3, the rate of emission reduction will surpass the rate of exposure increase. For example, when the restriction ratio is 0.4, the increase in exposure (7.4%) is less than the decrease in emissions (8%). When the restriction ratio exceeds 0.5, the emission reduction exceeds 11%. In regions where traffic emissions account for a significant portion of environmental pollution, the substantial reduction in traffic emissions will help transform a heavy pollution background into a light pollution background. Therefore, as shown in Figure 6(a), both exposure and emissions are expected to improve in a more favorable direction.

2. **Under bus fare changes**

Changes in bus fares have few impacts on modal choice, and both system emissions and exposure remain almost stable in LPS and HPS. Therefore, adjusting bus fares has not significantly altered emission and current exposure levels.

3. **Under ridesharing fare changes**

High ridesharing fares not only fail to reduce current travel exposure but also do not effectively reduce emissions. This effect is particularly significant under LPS. As shown in Figure 6(c) and Figure 6(d), as the ridesharing fare increases from 1 CNY/km to 10 CNY/km, exposure and emissions increase by 9.2% and 6.3%, respectively, which is notably higher than the increase under HPS (0.6% and 2.6%). Meanwhile, it is worth noting that under HPS, exposure initially slightly decreases with the fares rising, but rebounds after the fare reaches its minimum at a fare of 5 CNY/km, as depicted in Figure 6(d). However, emissions remain lowest at 1 CNY /km. Additionally, when ridesharing fares are high (such as ≥8 CNY/km under LPS or ≥9 yuan/km under HPS), both emissions and exposure grow faster. This illustrates that high ridesharing fares would exacerbate the current travel exposure and have an adverse effect on long-term urban traffic pollution management.

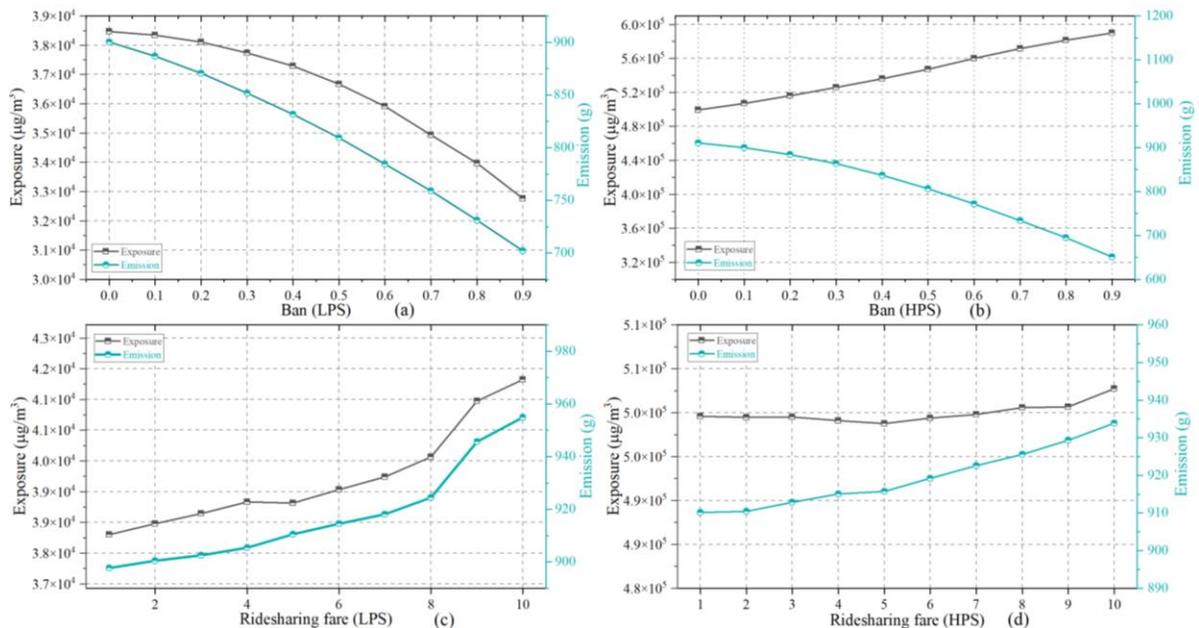

**Figure 6 System emissions and exposure under traffic measures**

*4.2.3. Time and monetary cost*

This section discusses changes in the average in-vehicle time, travel time, and monetary cost of the MT system.

1. **Under vehicle restriction**

As shown in **Figure 7** (a), under LPS, with the increase in restriction ratio, a slight decrease in in-vehicle time, indicating a reduction in road congestion. However, travel time gradually increases. At the same time, monetary cost also shows an upward trend, but the change is relatively small. However, the increase in monetary cost is more significant under HPS, especially when the vehcile restriction ratio is high, where the cost nearly grows exponentially, increasing the economic burden of the residents.

2. **Under bus fare changes**

Within the bus fare range of 1-10 CNY, whether under LPS or HPS, the in-vehicle time, travel time, and monetary cost stay almost stable, as shown in Figure 7(b). This indicates that within this fare range, changes in bus fares have a limited impact on these indicators.

3. **Under ridesharing fare changes**

when the ridesharing fare varies between 1-10 CNY/km, in-vehicle time, travel time, and monetary cost all show an increasing trend under LPS, as displayed in Figure 7 (c). Specifically, when the ridesharing fare exceeds 7 CNY/km, the increase in monetary cost becomes more significant. In contrast, Figure 7 (c) shows that, under HPS, monetary cost first decreases and then stabilizes, which is caused by more and more travelers choosing the low-exposure option of car travel.

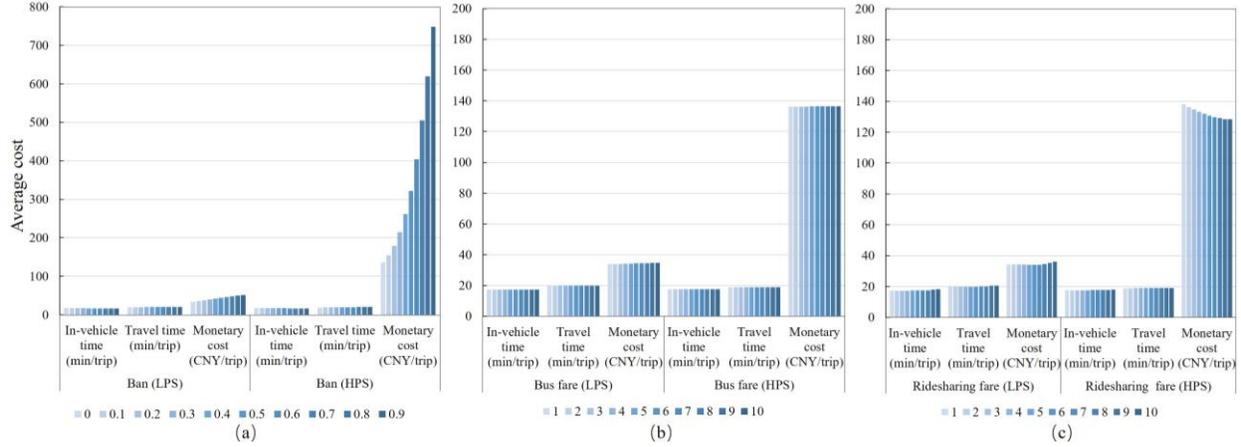

**Figure 7** Time and monetary cost under traffic measures

### 4.3. *Impact of exposure perception*

This section analyzes the impact of exposure perception $\omega$, where $\omega \in [0,0.8]$. According to the assumption in this paper, the higher the exposure perception of travelers, the more likely they are to adjust their travel behavior to reduce exposure risks. The experimental results show that as $\omega$ increases, when the perception is low, the system performance alters significantly. However, when the perception reaches a certain level, the impact on the system gradually weakens. This might be due to other factors such as time cost, which cause travelers with higher exposure perception to still tend to maintain their original travel decisions.

#### 4.3.1. *Modal choice*

As shown in Figure 8(a), in LPS, the impact of exposure perception on modal choice is relatively small. Although the increase in $\omega$ results in a higher proportion of solo and ridesharing, the use of public transport decreases accordingly, the change in modal split is minor. For example, compared to when $\omega = 0$, at $\omega = 0.8$, the ridesharing proportion increased by only 2%, while the public transportation proportion decreased by 4%. Meanwhile, the overall changes are concentrated at lower $\omega$ values.

Figure 8(b) presents the results in HPS. on the one hand, the overall changes are concentrated at lower $\omega$ values. For example, when $\omega$ is between 0 and 0.3, the changes in ridesharing and public transportation are more significant. However, as the exposure perception increases further, the changes in ridesharing and public transportation level off. On the other hand, heavier pollution exacerbates the variation in mode choice across different levels of exposure perception. As the perception increases, the proportion of ridesharing rises significantly, while the share of public transport declines markedly. Notably, when $\omega = 0.8$, the proportion of ridesharing exceeds that of public transport by 31%, a substantial increase compared to the 10% difference observed under LPS.

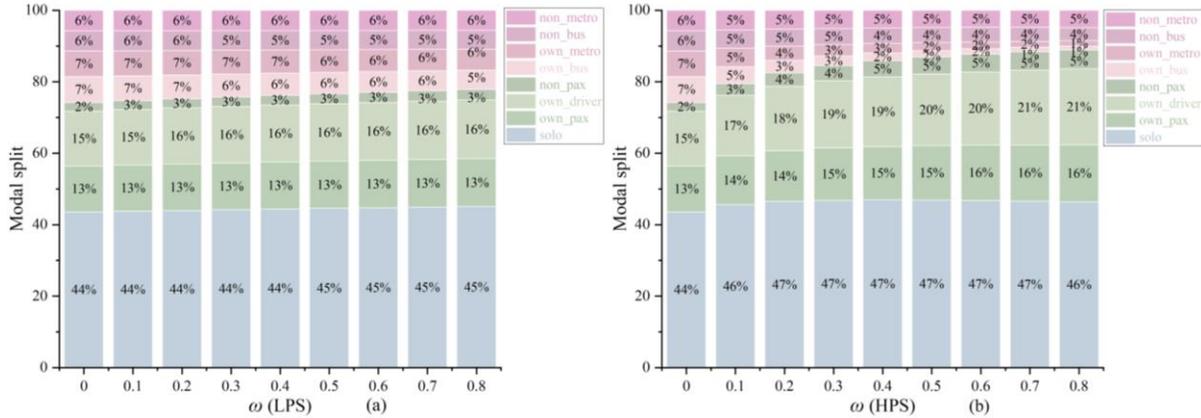

**Figure 8   Modal split under different values of exposure perception**

*4.3.2. Emission and exposure*

As exposure perception increases, total exposure in the system decreases under both light and heavy pollution scenarios, as shown in Figure 9. However, the emission trends vary between different scenarios. Under LPS, depicted in Figure 9(a), emission gradually increases with the rise in exposure perception. This trend is primarily driven by the increase in solo and ridesharing trips, leading to higher emission, which is unfavorable for long-term development. In contrast, under HPS, shown in Figure 9(b), emission reaches a turning point at $\omega = 0.2$, after which it begins to decline slowly. This change is likely attributed to the significant increase in ridesharing, which helps reduce emissions compared to the shift from public transport to solo travel. Although emissions at $\omega = 0.8$ are still 1.2% higher than at $\omega = 0$, the current exposure has decreased by 18.2%.

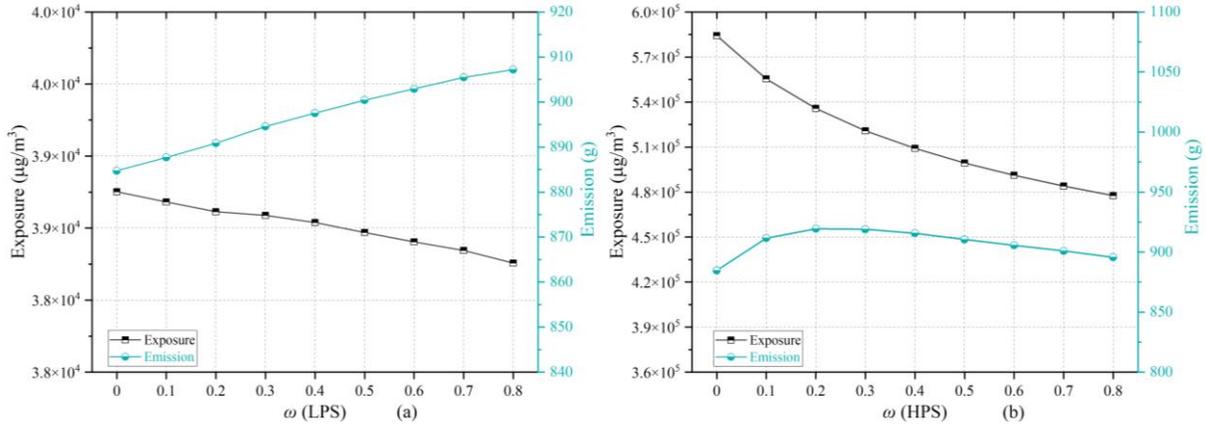

**Figure 9  Emission and exposure under different values of** $\omega$

*4.3.3. Monetary cost*

Figure 10 reveals that as exposure perception increases, monetary costs of the system also rise. When exposure perception is low, the growth in monetary cost is relatively rapid. For instance, compared to $\omega = 0$, when $\omega$ reaches 0.3, monetary costs have doubled under LPS, and increased by 7 times under HPS. As exposure perception continues to rise, monetary costs maintain an upward trend, but the rate of increase slows down compared to the low-perception case.

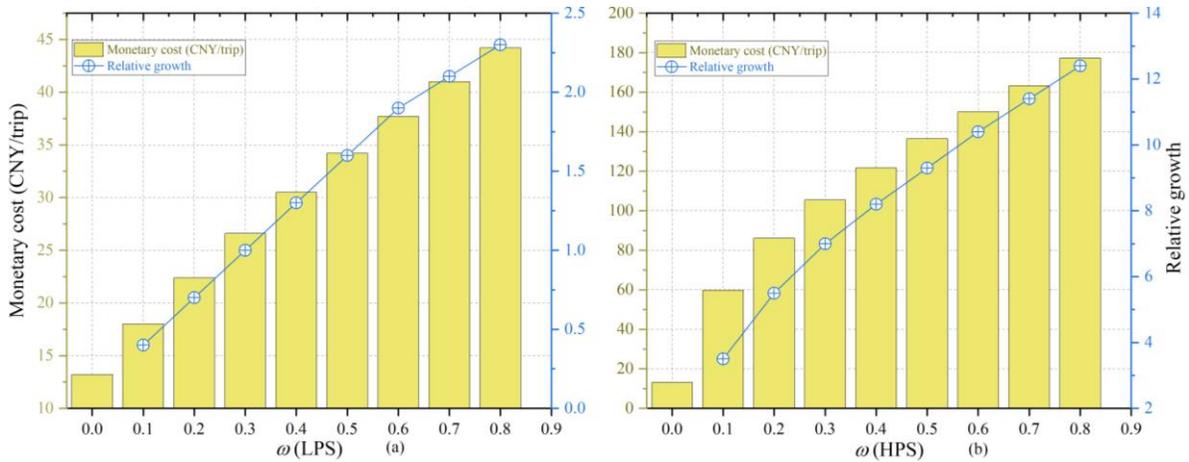

**Figure 10  Monetary cost changes under different values of** $\omega$

## 5. Policy implication and practice

This section, based on the experimental results from Section 4, presents policy suggestions from two perspectives: routine traffic management (under LPS) and emergency traffic management (under HPS), considering four dimensions—behavior, emissions, exposure, and costs.

## 5.1. Routine traffic management (RTM)

Under routine traffic management, with generally favorable air quality, the exposure risk is relatively low, requiring only moderate regulation of travel exposure.The primary goal should focus on meeting transportation demand while preventing the air quality from deteriorating into heavy pollution.

**(1) Vehcile restriction in RTM**

Under LPS, vehicle restriction policies effectively reduce car usage and increase the demand for public transport, thus facilitating a shift from cars to green transportation modes, reducing traffic emissions and exposure levels. However, such restrictions would slightly increase monetary costs and impose some economic pressure. Therefore,

- Implement gradually: Gradually adjust the intensity of restrictions and avoid excessively high restriction ratios to avoide heavy socio-economic burdens.
- Promote green travel: Combine vehicle restriction policies with improvements in public transportation services to meet travel demands while encouraging travelers to choose efficient and environmentally friendly transportation modes (e.g., buses and metro).

**(2) Pricing measures in RTM**

In routine management, attention should be given to reducing ridesharing and public transport costs.

- Optimize bus fares: Increasing bus fares would cause travelers to shift towards cars (including solo and ridesharing), which negatively impacts the reduction of traffic emissions and the promotion of green travel. However, compared to other measures, the impact of fare changes on the system is relatively small. Therefore, bus fare strategies should be optimized, combining fare discounts with improved service quality to enhance the attractiveness of public transport.
- Regulate ridesharing fares: Higher ridesharing prices lead to an increased proportion of solo car travel, as both ridesharing and public transport trips shift to solo driving. This results in higher emissions and exposure levels. To address this, providing subsidies to the ridesharing market can lower fees and stimulate ridesharing demand, thereby reducing solo car usage and mitigating both emissions and travel exposure.

**(3) Information guidance in RTM**

Increased exposure perception encourages car use. Under LPS, while travel exposure decreases, emissions continue to rise, potentially worsening air quality and leading to heavy pollution. Therefore, in routine management (including the transition from emergency status to normal conditions), information guidance is important.

- Update air quality information timely: Disseminate details about the favorable air quality conditions under LPS, particularly the transition from an emergency to routine state. It contributes to guide travelers towards public transport or other clean transportation modes, avoiding the shift towards polluting modes due to overestimated risks. This will help promote green and sustainable develoment, especially when exposure perception is high.

- Enhance information transmission effectiveness: Use various formats, such as visualization and time-based segmentation, to enhance the accuracy and reach of the information.

### 5.2. *Emergency traffic management (ETM)*

Air quality worsens, increasing health risks. Affected by heavy air pullution, emergency traffic management should focus not only on travel demand but also on reducing travel exposure. Section 4 reveals significant differences in the effectiveness of management measures between LPS and HPS. Thus, it is crucial to examine how worsening pollution interacts with measures to inform optimal urban traffic control strategies.

**(1) Vehicle restriction in ETM**

Under heavy pollution, traffic restrictions can reduce emissions significantly but temporarily increase exposure. However, when the restriction rate is appropriate (e.g., 30%), the reduction in emissions outweighs the rise in exposure risk.

- Optimize the restriction ratio: Given the major role of traffic emissions in pollution, implementing restrictions is expected to shift the scenario from heavy to light pollution, leading to long-term improvements in both exposure and emissions. It is suggested choose a moderate restriction ratio (e.g., 30%-50%) to reduce emissions while controlling the growth of monetary costs and exposure risks.

- Improve high-exposure environments: Improve ventilation and protective measures at high-exposure locations, such as bus stations, and advise the public to wear masks to reduce health risks.

**(2) Pricing measures in ETM**

Under HPS, the impact of bus fares is almost negligible, and ridesharing prices need further optimization in emergency management.

- Synergize bus fare with other measures: Changes in bus fares have a negligible impact on mode choice, and the indicators related to emissions, time, money, and exposure generally remain unchanged. However, combining fare reductions with traffic restrictions can further enhance the overall effectiveness of emergency measures.

- Ridesharing pricing: The combined effect of heavy pollution and higher ridesharing fees creates a non-monotonic pattern in exposure, initially decreasing before rising, while emissions continue to increase. Emissions are lowest when ridesharing fees are lower. These varying effects on exposure and emissions highlight the need for further research to determine an optimal ridesharing fee structure that diminish current exposure risks and magnify long-term environmental benefits.

**(3) Information dissemination in ETM**

During heavy pollution, increasing exposure perception (e.g., making $\omega > 0.2$) can decrease both emissions and exposure levels in the system. Therefore:

- **Issue pollution alerts:** heavy pollution alert mechanism can enhance the public's awareness of pollution levels, coupled with protective reminders (such as reducing travel and wearing masks, to help travelers access information about air pollution and assess their exposure, improving their exposure perception.

- **Balance exposure perception:** Ecessively high exposure perception could lead to increased monetary costs and excessive panic. It is crucial to ensure the accuracy of air quality monitoring, provide tiered warnings, and offer clear, actionable recommendations to prevent unnecessary anxiety and economic strain.

## 6. Conclusion and future study

This study investigates the complex relationship between air pollution exposure and MT systems. By developing a model that integrates air pollution exposure and using numerical experiments, we analyze the system's response to the disturbances, like air pollution, traffic measures, and exposure perception, showing that these factors significantly affect travel behavior, urban environments, and public health in MT system. It provides theoretical and technical support for modeling MT system under urban air pollution governance. Particularly, it proposes policy recommendations for both routine and emergency traffic management, enhancing the resilience of transportation and supporting the sustainable development of cities.

The findings show that the effectiveness of traffic-related management measures varies significantly under different pollution levels, especially between light and heavy pollution scenarios. Vehicle restrictions effectively reduce car use and promote greener travel modes such as public transport and ridesharing, lowering both emissions and exposure in light pollution. Additionally, lower bus fare and ridesharing fee are beneficial for both emission reduction and exposure reduction. However, under heavy pollution, more precise adjustments to restriction ratios are needed, as excessive restrictions would increase exposure and economic costs. Meanwhile, changes in the MT system due to ridesharing fare variations are more complex, requiring further exploration and consideration of trade-offs. It is important to combine measures like vehicle restrictions, bus fare reduction, and ridesharing fare exploring under heavy pollution, to balance emission reductions and exposure control, ensuring efficient travel while reducing air pollution risks.

Future research could focus on ridesharing pricing under heavy pollution, the joint implementation of multiple measures, and the interaction between traffic policy and exposure perception. These investigations will improve MT models and provide practical solutions for sustainable urban transportation. Additionally, we plan to refine the model using real road networks and OD data and conduct surveys to better understand exposure perception across cities, improving smart management strategies for air pollution exposure.